# A Minimax Bias Estimator for OLS Variances under Heteroskedasticity

Mumtaz Ahmed[1] and Asad Zaman[2]

**May 2014**

**Abstract**

Analytic evaluation of heteroskedasticity consistent covariance matrix estimates (HCCME) is difficult because of the complexity of the formulae currently available. We obtain new analytic formulae for the bias of a class of estimators of the covariance matrix of OLS in a standard linear regression model. These formulae provide substantial insight into the properties and performance characteristics of these estimators. In particular, we find a new estimator which minimizes the maximum possible bias and improves substantially on the standard Eicker-White estimate.

---

[1]Corresponding author, Assistant Professor, Department of Management Sciences, COMSATS Institute of Information Technology, Islamabad, Pakistan, Email: mumtaz.ahmed@comsats.edu.pk, mumtaz.mumtazahmed@gmail.com

[2]Vice Chancellor, Pakistan Institute of Development Economics, Islamabad, Pakistan. Email: asadzaman@alum.mit.edu

## 1. Introduction

Assessing whether or not a regressor affects the dependent variable is often an important concern in a regression model. Statistical significance is judged by looking at the standard error of the estimated coefficient. This standard error can be inaccurate, and lead to wrong decisions regarding significance, when heteroskedasticity is ignored. Eicker's (1963) discovery that one can obtain consistent estimates of these standard errors, even with an infinite dimensional nuisance parameter, was a breakthrough. White's (1980) generalization of the idea to dynamic models has spawned an extensive literature. The initial proposals of Eicker-White were found to have rather large small sample biases (see for example, Chesher and Jewitt (1987)). A number of alternatives with reduced biases have been proposed; for example, Hinkley (1977), Horn et al. (1975), Mackinnon and White (1985), Cribari-Neto (2004) and Cribari-Neto et al. (2007).

This paper provides an analytic evaluation of the biases of a class of estimators which includes the original Eicker-White estimator and some of its suggested bias-corrected versions. Since complexity of the algebra has hindered analytic evaluation, simulation studies have been used for such evaluations in the past. However bias depends on both the sequence of regressors and the sequence of variances; this is an extremely high dimensional space which cannot be explored adequately via simulations. We develop analytic formulae for the bias, and show that these can be used to provide global and analytic evaluations of biases. Since the heteroskedastic sequence of variances is typically unknown, we compute the least favorable configuration of heteroskedasticity, which maximizes the bias. Within a one parameter family which embeds the original Eicker-White estimator, we compute an estimator which minimizes this maximum. Our bias formulae show that this minimax estimator improves substantially on Eicker-White in terms of bias.

## 2. The Framework

In this section, we set out the basic model and definitions required to state our results. Consider a linear regression model with y=Xβ+ε, where y is a T x 1 vector of observations on the dependent variable, X is a T x K matrix of regressors, β is a K x 1 vector of parameters, and ε is a T x 1 vector of errors. We allow for heteroskedasticity by assuming that ε is N(0,Σ) where Σ is a diagonal matrix with $\Sigma_{tt}=\sigma_t^2$ . The OLS estimate of the coefficient $\beta$ is $\hat{\beta} = (X'X)^{-1}X'y$.

The covariance matrix of OLS estimates of $\beta$ is $\Omega = (X'X)^{-1}X'\Sigma X(X'X)^{-1}$. It is convenient to initially consider the case of a single regressor, with K=2. Thus we assume that the first column of X consists of 1's, while the second column is the sole regressor $x = (x_1, x_2, \ldots, x_T)'$. The algebra simplifies substantially upon assuming that $\sum_{t=1}^{T} x_t = 0$. This can be assumed without loss of generality by taking differences of the regressors from the mean. Extensions of our results to the case of multiple regressors are discussed in the concluding section of the paper.

Our goal in this paper is to derive analytical expressions for the bias of a class of heteroskedasticity consistent estimators of $\Omega_{22}$, the variance of $\hat{\beta}_2$ under heteroskedasticity. These permit us to calculate an estimator which minimizes the maximum bias, and improves substantially on the Eicker-White estimator.

Both the derivation and interpretation of our results to follow become much simpler using the following notational device. Let $s^2 = \frac{1}{T}\sum_{t=1}^{T} x_t^2$ be the second moment of regressor $x$; note that $s$ is the Standard Deviation since the regressors are centered. Introduce artificial random variables $(Z, V)$, which take T possible values $\left(\frac{x_t}{s}, \sigma_t^2\right)$ for $t = 1, 2, \ldots T$, each with probability $(1/T)$. Let $z_t = \frac{x_t}{s}, t = 1, 2, \ldots, T$. Since regressors are centered around their mean, $EZ = \frac{1}{T}\sum_{t=1}^{T} z_t = \frac{1}{T}\sum_{t=1}^{T}\frac{x_t}{s} = 0$. In addition, note that, $EZ^2 = 1$, since $s^2 = \frac{1}{T}\sum_{t=1}^{T} x_t^2$.

**Lemma 1**: The variance of the OLS estimate $\hat{\beta}_2$ with heteroskedasticity is:

[1]
$$\Omega_{22} = \frac{Cov(VZ,Z)}{Ts^2}$$

**Proof**: Note that $X'X = T\begin{bmatrix} 1 & 0 \\ 0 & s^2 \end{bmatrix}$ and (i,j) entry of $X'\Sigma X$ is $TsEVZ$. The diagonal entries of $\Omega$ corresponding to regressor Z is $\Omega_{22} = \frac{EVZ^2}{Ts^2}$. This proves the lemma. ∎

Note that, since $Cov(V,Z^2)=EVZ^2 - (EV)(EZ^2)$, we can write $\Omega_{22} = \frac{EVZ^2}{Ts^2} = \frac{Cov(V,Z^2)}{Ts^2} + \frac{EV}{Ts^2}$. This implies that the formulae for $\Omega$ under homoskedasticity and heteroskedasticity coincide when $Cov(V, Z^2) = 0$. This means that the usual estimates of the variance of $\hat{\beta}_2$ will be unbiased and consistent under this condition, that the sequence of variances is not correlated with the squared sequence of regressors.

## 3. Bias of EW-type Estimates

We will now derive analytical expressions for the bias of a class of estimators which includes the Eicker-White, as well as the Hinkley bias-corrected version of the heteroskedasticity consistent covariance matrix estimator (HCCME).

From Lemma 1, the true variance of the OLS estimate $\hat{\beta}_2$ of $\beta_2$, is given by:

[2]
$$\Omega_{22} = \frac{Cov(VZ,Z)}{Ts^2} = \frac{EVZ^2}{Ts^2} = \frac{1}{T^2 s^2} \sum_{t=1}^{T} z_t^2 \sigma_t^2$$

The class of estimators under consideration replaces the variances by the scaled squares of the corresponding residuals – $(1 + a/T)$ being the scaling factor:

[3]
$$\hat{\Omega}_{22}(a) = \frac{1}{T^2 s^2}\left(1+\frac{a}{T}\right)\sum_{t=1}^{T} z_t^2 e_t^2$$

Though the squares of the OLS residuals are approximately unbiased for the corresponding error variances in large samples, these are not consistent estimators for these variances; we have only one observation per variance, and this does not change with increasing sample size. Nonetheless, as Eicker (1963) showed, the weighted sum provides a consistent estimator for the variance of the OLS estimate of $\beta_2$ under heteroskedasticity. The following theorem computes the bias of this class of estimators.

**Theorem 1:**

The bias $B_{22}(a) = E\hat{\Omega}_{22}(a) - \Omega_{22}$ of the estimator $\hat{\Omega}_{22}(a)$ for the variance of $\beta_2$ is:

[4] $B_{22}(a) = \frac{1}{T^3 s^2} \sum_{t=1}^{T}\left[1+\frac{a}{T}+2\left(1+\frac{a}{T}\right)(EZ^3)z_t + \left\{a+\left(1+\frac{a}{T}\right)(EZ^4-2)\right\}z_t^2 - 2\left(1+\frac{a}{T}\right)z_t^4\right]\sigma_t^2$

**Proof:** From the expressions for $\hat{\Omega}_{22}(a)$ and $\Omega_{22}$ given earlier, we get:

[5] $$B_{22}(a) = E\hat{\Omega}_{22}(a) - \Omega_{22} = \frac{1}{T^2 s^2} \sum_{t=1}^{T} z_t^2 \left[ \left(1 + \frac{a}{T}\right) E(e_t^2) - \sigma_t^2 \right]$$

To compute this, we need $E(e_t^2)$, which is given by following lemma:

**Lemma 2:** The expected value of OLS squared residuals is given by:

[6] $$E(e_t^2) = \sigma_t^2 + \frac{1}{T}\left\{ EV - 2\sigma_t^2 - 2z_t^2 \sigma_t^2 + 2z_t EVZ + z_t^2 EVZ^2 \right\}$$

**Proof of Lemma 2:** The OLS residuals are $e = y - X\hat{\beta} = (I - H)\varepsilon$, where, $H = X(X'X)^{-1}X'$ is the 'hat matrix' as usual. The (s,t) entry of the T×T matrix H is: $H_{st} = \frac{1}{T}(1 + z_s z_t)$.

Now note that $e_t = \varepsilon_t - \sum_{i=1}^{T} h_{ti} \varepsilon_i = (1 - h_{tt})\varepsilon_t - \sum_{\substack{i=1 \\ i \neq t}}^{T} h_{ti} \varepsilon_i$.

Since $E\, e=0$, and the $\varepsilon$'s are independent, the variance of e is the sum of the variances. This can be explicitly calculated as follows:

$$E(e_t^2) = (1 - h_{tt})^2 \sigma_t^2 + \sum_{i=1, i \neq t}^{T} h_{ti}^2 \sigma_i^2 = (1 - 2h_{tt})\sigma_t^2 + \sum_{i=1}^{T} h_{ti}^2 \sigma_i^2$$

$$= \left(1 - \frac{2}{T}(1 + z_t^2)\right)\sigma_t^2 + \sum_{i=1}^{T} \left(\frac{1}{T}(1 + z_t z_i)\right)^2 \sigma_i^2$$

Noting that $EVZ^j = (1/T) \sum z_t^j \sigma_t^2$, it follows that:

$$E(e_t^2) = \sigma_t^2 - \frac{2}{T}\sigma_t^2 - \frac{2}{T} z_t^2 \sigma_t^2 + \frac{1}{T} EV + \frac{1}{T} z_t^2 EVZ^2 + \frac{2}{T} z_t EVZ$$

This is easily translated into the expression given in the Lemma. ∎

Substituting the expression of the Lemma into equation [5] above leads to the expression given in Theorem 1. ∎

## 4. *Maximum Bias*

Having analytical expressions for the bias allow us to calculate the configuration of variances which leads to the maximum bias. In this section we characterize this least favorable form of heteroskedasticity, and the associated maximum bias. We first re-write the expression for bias in a form that permits easy calculations of the required maxima.

Define polynomial, $p(\lambda)$, as

$$[7] \quad p(\lambda) = 1 + \frac{a}{T} + 2\left(1 + \frac{a}{T}\right)(EZ^3)\lambda + \left\{a + \left(1 + \frac{a}{T}\right)(EZ^4 - 2)\right\}\lambda^2 - 2\left(1 + \frac{a}{T}\right)\lambda^4$$

From the expression for bias given in Theorem 1 of the previous section, we find that

$$[8] \quad B_{22}(a) = \frac{1}{T^3 s^2} \sum_{t=1}^{T} p(z_t) \sigma_t^2$$

If the variances are unconstrained, then the bias can be arbitrarily large, so we assume some upper bound on the variances: $\forall t : \sigma_t^2 \leq U$. These variances are unknown, unobservable, and cannot be estimated. Thus we focus on finding the least favorable configuration of variances, the one which gives the maximum possible bias. It turns out that maximizing positive and negative biases requires separate least favorable configurations, so we calculate the two separately.

**Theorem 2:** Let $\mathcal{B}_i^+$ and $\mathcal{B}_i^-$ be the maximum possible positive and negative biases of the EW-type estimators $\hat{\Omega}_{22}(a)$ (defined in equation [3] above) for the variance of $\beta_2$. These are given by:

$$[9] \quad \mathcal{B}^+ = \max_{\sigma_t^2 \leq U} B_{22} = \frac{1}{T^3 s^2} \sum_{t=1}^{T} \max(p(z_t), 0) U$$

[10] $$\mathcal{B}^- = \min_{\sigma_t^2 \leq U} B_{22} = \frac{1}{T^3 s^2} \sum_{t=1}^{T} \min(p(z_t), 0) U$$

**Proof:** This is trivial, since we are maximizing a sum of linear functions. Each term in the sum can be maximized separately with respect to variances ($\sigma_t^2$). The positive maximum occurs when $\sigma_t^2$ is set to its maximum possible value $U$ when the coefficient is positive, and its minimum possible value '0' when the coefficient is negative. Reversing these choices gives the largest negative bias. ∎

### 4.1. An Example: Asymptotic Calculations with Normal Regressor

To provide a concrete illustration of the use of these formulae, we do more explicit asymptotic calculations for the case of normal regressors. We assume that the regressors are drawn from an i.i.d. N(0,1) sequence. In this case, X is a N(0,1) random variable, i.e. $EX = 0$ and $s^2 = Var(X) = EX^2 = 1$. Note that, here, Z=X, i.e. EZ=0 and EZ$^2$=1. In large samples, the skewness $EZ^3$ should be approximately zero, while the kurtosis, $EZ^4$ should be approximately '3'. Making these asymptotic approximations, the polynomial $p(\lambda)$ simplifies to:

[11] $$p(\lambda) = 1 + \frac{a}{T} + \left(a + 1 + \frac{a}{T}\right)\lambda^2 - 2\left(1 + \frac{a}{T}\right)\lambda^4$$

Asymptotically, $p(\lambda)$ is a quadratic in $\lambda^2$ with a unique positive root:

[12] $$r = \frac{1 + a + \sqrt{8 + (1+a)^2}}{4}$$

Because the coefficient of $\lambda^4$ is negative, $p(z) > 0$ if and only if $z^2 < r$. This allows explicit asymptotic evaluation of the expressions for bias given in Theorem 2 above. Let $\phi$ and $\Phi$ be the density and cumulative distribution function of a standard normal random variable. Taking the limit as 'T' goes to infinity, the maximum positive and negative biases of the estimator $\hat{\Omega}_{22}(a)$ can be written as:

[13] $$\mathcal{B}^+(a) = 2\{2r - a + 5\}\sqrt{r}\,\phi(\sqrt{r}) + 2(a-4)\Phi(\sqrt{r}) - a + 4$$

[14] $$-B^-(a) = 2\{2r - a + 5\}\sqrt{r}\,\phi(\sqrt{r}) + 2(a-4)\Phi(\sqrt{r}) - 2a + 8$$

For details of these calculations, see Ahmed (2012). These bias functions are plotted in Figure 1 below. The maximum bias is the larger of the two, and the minimax value of '$a$' is the one at which the two functions are equal. It is easily calculated that $a = 4$ is the minimax value for '$a$'. The original Eicker-White estimator ($a = 0$) has a maximum risk of 4.66, while the Hinkley correction with $a = 2$, has a maximum risk of 3.96. The minimax estimator with $a = 4$, improves on both with '$a$' maximum risk of 3.67.

**Figure 1: Maximum Positive and Negative Biases of HCCME by Varying '$a$'**

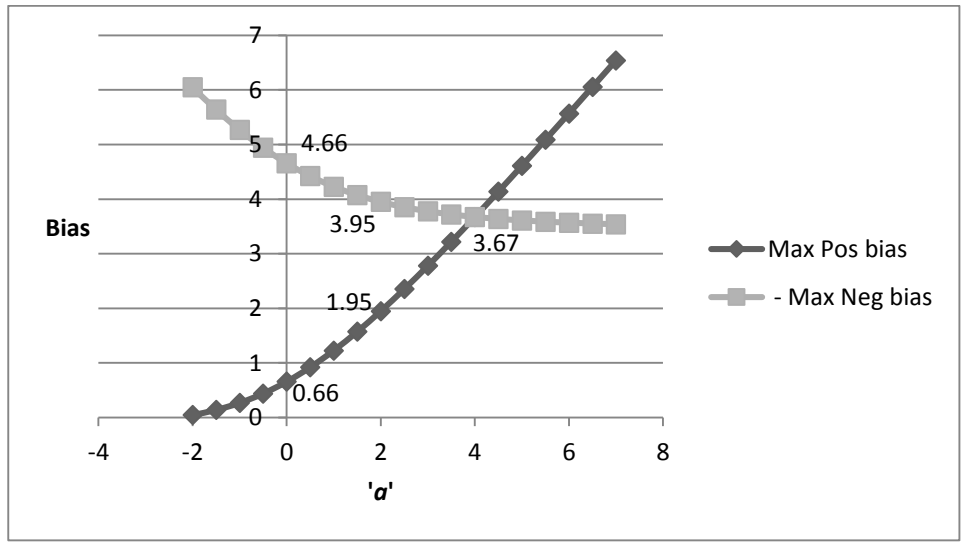

**Note: Max Pos and Max Neg denote the maximum positive and maximum negative biases respectively.**

## 5. *The Minimax Bias Estimator – A Computable Case*

We now construct a special sequence of regressors for which an exact analytic computation of the minimax bias estimator is possible. Specifically, this sequence allows us to analytically compute the bias functions and hence the minimax value of '$a$'.

Choose sample size '$T$' and constant $M > 1$ such that $k = T/(2M^2)$ is an integer, and consider the sequence of regressors $x_1, x_2, \ldots, x_T$ such that $x_1 = x_2 = \cdots = x_k = -M$, $x_{k+1} = \cdots = x_{T-k} = 0$, and $x_{T-k+1} = \cdots = x_T = +M$. Let Z be the random variable such that $z_t = x_t$ with probability 1/T; note that Z takes just three possible values –M, 0 and M. We can easily check that $EZ = 0$, $EZ^2 = 1$, $EZ^3 = 0$ and $EZ^4 = M^2$. That is Z is centered and standardized and has zero skewness and with kurtosis $K = EZ^4 = M^2$. So, we can write the polynomial $p(z)$ as:

[15] $$p(z) = \left(1 + \frac{a}{T}\right) + \left(a + \left(1 + \frac{a}{T}\right)(M^2 - 2)\right)z^2 - 2\left(1 + \frac{a}{T}\right)z^4$$

Note that $p(0) = 1 + a/T > 0$ for all positive values of 'a'. Further note that, $p(+M) = p(-M)$ is a polynomial in $M^2$ with positive root:

[16] $$r^+ = \frac{\left(a - 2\left(1 + \frac{a}{T}\right)\right) + \sqrt{\left(a - 2\left(1 + \frac{a}{T}\right)\right)^2 + 4\left(1 + \frac{a}{T}\right)^2}}{2\left(1 + \frac{a}{T}\right)}.$$

Note that $p(\pm M) < 0$, $\forall M^2 > r^+ \geq 1$ and $p(\pm M) > 0$, $\forall 1 < M^2 < r^+$

First we consider the case where the value of 'a' is below $M^2 + 2 - \frac{1}{M^2}$, and T is large. In this case, it is easily checked from above calculations that $p(\pm M) < 0$. The maximum positive and negative bias functions can be calculated as follows:

$$\mathcal{B}^+ = \frac{1}{T^3 s^2} \sum_{t=k+1}^{t=T-k} p(z_t) U = \frac{1}{T^2 s^2} \frac{T - 2k}{T} p(0) U = \frac{1}{T^2 s^2}\left(1 - \frac{1}{M^2}\right)\left(1 + \frac{a}{T}\right) U$$

$$\mathcal{B}^- = \frac{1}{T^3 s^2}\left\{\sum_{t=1}^{k} p(z_t) U + \sum_{t=T-k+1}^{T} p(z_t) U\right\} = \frac{1}{T^2 s^2} \frac{2k}{T} p(M) U = \frac{1}{T^2 s^2} \frac{p(M) U}{M^2}$$

$$= \frac{1}{T^2 s^2} \frac{U}{M^2}\left\{\left(1 + \frac{a}{T}\right) + \left(a + (1 + a/T)(M^2 - 2)\right)M^2 - 2\left(1 + \frac{a}{T}\right)M^4\right\}$$

Simplifying, the above expressions, we get:

[17] $$M^2\{T^2 s^2\}\mathcal{B}^+/U = (M^2 - 1) + \frac{a}{T}(M^2 - 1)$$

[18] $$-M^2\{T^2 s^2\}\mathcal{B}^-/U = M^4 + (2 - a)M^2 - 1 + \frac{a}{T}(M^4 + 2M^2 - 1)$$

Note that, here $s^2 = 1$, so we have,

[19] $$M^2 T^2 \mathcal{B}^+/U = (M^2-1) + \frac{a}{T}(M^2-1)$$

[20] $$-M^2 T^2 \mathcal{B}^-/U = M^4 + (2-a)M^2 - 1 + \frac{a}{T}(M^4 + 2M^2 - 1)$$

Solving $M^2 T^2 \mathcal{B}^+/U = -M^2 T^2 \mathcal{B}^-/U$ yields the minimax (optimal) value of 'a' in case of finite samples:

[21] $$a^* = \frac{M^2+1}{1-\frac{1}{T}(M^2+1)} = \frac{K+1}{1-\frac{1}{T}(K+1)}$$

This proves that $a^*$ minimizes the maximum bias for the range of values of $a < M^2 + 2 - \frac{1}{M^2}$. Note that this range includes $a=0$ and $a=2$, so that the minimax estimator $a^*$ dominates Eicker-White and Hinkley in terms of maximum risk. For larger values of 'a', the polynomial $p(z)$ becomes positive at $+M$ and $-M$ so that the above calculations do not apply. However, it is easily checked that the maximum risk becomes larger than those given by the formulae above, so that larger values of 'a' cannot lead to reductions in maximum risk. It follows that this value of '$a^*$' minimizes the maximum risk over all possible non-negative values of 'a'. This completes the results for finite samples.

Taking the limit at T goes to infinity, we can get the asymptotic minimax value of 'a', which is given below:

[22] $$a^* = M^2 + 1 = K + 1.$$

Note that when K=3, matching the kurtosis of normal regressors, we get the same asymptotic minimax value of $a=4$.

## 6. Generalization to Arbitrary Sequence of Regressors

The special case discussed in the previous section was designed to provide a base analytical formula for the minimax value of 'a'. While analytics are not possible for a general sequence of regressors, numerical calculations are straightforward. For any fixed sequence of regressors, we can numerically compute the positive and negative bias functions, and the value of '$a$' at the point of their intersection – which is the minimax value of '$a$'. Our goal was to use the analytical formula as a leading term for '$a$' and get a better fit for the general case by using a response surface Monte Carlo study.

However, numerical calculations led to the surprising conclusion that the minimax value of '$a$' depends only on the sample size 'T' and the kurtosis 'K' of the sequence of regressors. Due to the complexity of the relation between regressors and the minimax value of '$a$', we were unable to establish this result analytically. The conjecture that only $K$ and T determine '$a$' is also supported by the illustrative example of the normal regressors, which leads to exactly the same asymptotic minimax value of '$a$' as this rather different sequence of discrete regressors. If this invariance conjecture is valid, this would lead to:

**(Conjectured) Theorem 3:** Assume that the variances are bounded so that $\forall t : \sigma_t^2 \leq U$. For each $a$, define the maximum bias $MB(a) = \max_{\sigma_t^2 \leq M} B_{22}(a)$ -- this is maximum possible bias obtainable by setting the heteroskedastic sequence of variances to the least favorable configuration. Let $K = EZ^4$ be the kurtosis of the sequence of regressors. Define, $a^* = \dfrac{K+1}{1 - \dfrac{1}{T}(K+1)}$

Then $MB(a^*) \leq MB(a)$ for all $a$.

**Heuristic Proof:** This was done by simulations. For each sample size 'T', and fixed value of kurtosis '$K$', we generated random sequences of 'T' regressors having kurtosis '$K$', and numerically computed the two bias functions $\mathcal{B}^+$ and $\mathcal{B}^-$. Setting the two equal led to the minimax value of '$a$'. Our expectation was that this computed minimax value would vary according to other features of the regressors. In particular, other moments, like the skewness would affect this value. To our surprise, this minimax value always came out the same. Some details of simulation results is provided in the following table, where we numerically calculated maximum bias and the optimal value of '$a$', where both positive and negative bias are equal, for six different randomly generated sequences of regressors with matching kurtosis and skewness measures. In particular, we fixed kurtosis $K$ at '3' and took two values of skewness, first fixing it to '0' and then to '1' and calculated maximum bias and also calculated optimal value of '$a$' for all '12' samples (six samples with skewness '0' and the remaining six samples with skewness '1'). The second half of table shows results with kurtosis measure fixed at '4' while skewness measure takes values zero and one respectively.

**Table 1: Maximum Bias under Different Set of Regressors with Varying Skewness and Fixed Kurtosis (Sample size: 100)**

| Samples of Regressors | K=3 & S=0 | | K=3 & S=1 | | K=4 & S=0 | | K=4 & S=1 | |
|---|---|---|---|---|---|---|---|---|
| | $a^*$ | MB | $a^*$ | MB | $a^*$ | MB | $a^*$ | MB |
| 1 | 4.167 | 2.080 | 4.167 | 2.070 | 5.263 | 3.046 | 5.263 | 2.885 |
| 2 | 4.167 | 2.012 | 4.167 | 2.131 | 5.263 | 4.166 | 5.263 | 4.095 |
| 3 | 4.167 | 2.394 | 4.167 | 1.833 | 5.263 | 4.162 | 5.263 | 3.750 |
| 4 | 4.167 | 2.771 | 4.167 | 1.933 | 5.263 | 3.535 | 5.263 | 2.743 |
| 5 | 4.167 | 2.137 | 4.167 | 1.837 | 5.263 | 4.166 | 5.263 | 3.358 |
| 6 | 4.167 | 2.256 | 4.167 | 2.070 | 5.263 | 3.391 | 5.263 | 3.358 |

**Note:** K and S are Kurtosis and Skewness measures respectively and MB represents the maximum bias.

We can see from the above table that, for all six samples with randomly chosen regressors, whether skewness is zero or one but with same kurtosis measure, the optimal value of '$a$' is same. Similar results were obtained for sample sizes 25 and 50 which are not reported to save space.

It is worth noting that our contribution does not depend on the validity of our conjecture – for any given fixed sequence of regressors, we can numerically compute the minimax value of '$a$' using the formulae developed. The calculations that we did for randomly generated sequences of regressors all led to the value of '$a$' described in the Theorem 3 above. This leads us to believe that this is a general result valid for all sequences of regressors.

## 7. *Evaluation of Relative Performance*

We analytically compare the relative performance of the Eicker-White, Hinkley and the minimax estimator in terms of asymptotic bias formulae obtained in [19] and [20]. Note that both the maximum bias functions $\mathcal{B}^+$ and $\mathcal{B}^=$ are proportional to $U$, the upper bound on the variances. To get a reasonable performance measure which is invariant to this arbitrary upper bound, it seems reasonable to divide by this factor. The asymptotic maximum positive bias is then:

[23] $\qquad \lim_{T\to\infty}\{T\mathcal{B}^+/U\} = 1 - 1/M^2$

Note that this does not depend on '$a$', and is bounded above by '1'.

On the other hand, the asymptotic maximum negative bias is:

[24] $\qquad \lim_{T\to\infty}\{-T\mathcal{B}^-/U\} = M^2 + (2-a) - 1/M^2$

Note that, this increase with the kurtosis ($K = M^2$) of regressors and is unbounded.

Let maximum of both biases (maximum positive and minus the maximum negative bias) is represented by: $\mathcal{B} = \max(\mathcal{B}^+, -\mathcal{B}^-)$, where $\mathcal{B}^+$ and $\mathcal{B}^=$ respectively are positive and negative biases. For the Eicker-White estimator with $a = 0$, this maximum bias is, $\mathcal{B}^{EW} = K + 2 - \frac{1}{K}$ which is somewhat larger than the kurtosis ($K$). Hinkley's bias correction has $a = 2$, which yields the maximum bias of: $\mathcal{B}^{Hi} = K - \frac{1}{K}$. Note that, this bias correction is too timid – it knocks out the middle term, but does nothing to the dominant bias term $K = M^2$.

The minimax value sets $a = M^2 + 1$, which results in:

[25] $\qquad \lim_{T\to\infty}\{-T\mathcal{B}^-/U\} = 1 - 1/M^2 = \lim_{T\to\infty}\{T\mathcal{B}^+/U\}$

Thus maximum bias of the minimax estimate is: $\mathcal{B} = 1 - \frac{1}{K}$. By knocking out the leading bias term, this results in maximum bias bounded above by 1.

When Kurtosis ($K$) is large, the minimax estimator is substantially superior to both Eicker-White and Hinkley in terms of maximum possible bias.

## *8. Conclusions*

We have obtained elementary explicit analytical formulae for the bias of variance estimates in a simple linear regression model with one regressor under heteroskedasticity. This allows us to calculate the pattern of the least favorable heteroskedastic sequence, and to compute worst case bias. In the past, simulation studies chose different patterns of heteroskedasticity in an ad-hoc fashion. This ad-hoc choice does not allow for accurate evaluation of strengths and weaknesses of different classes of estimators. Our methodology permits an analytical assessment and comparison of estimators on the basis of their worst case bias.

Our results permit numerical calculation of the minimax estimator for any sequence of regressors. One important payoff from our research is an explicit formula for a minimax estimator which has substantially lower maximum bias than conventional estimators. We derived the analytic formula for a restricted class of regressor sequences. Our simulations show that this same formula remains valid for randomly chosen sequences of regressors. Assuming this invariance, our results show that a single regressor is significant at the 95% level under least favorable configurations of heteroskedasticity if zero is not contained in the interval $\hat{\beta}_2 \pm 2\left[\frac{1}{Ts^2}\sqrt{\left(1+\frac{K+1}{T}\right)\sum_{t=1}^{T}x_t^2 e_t^2}\right]$.

Research is under way on extensions to multiple regressor case; this requires substantially more complex algebra. However our current results can be used to provide bounds on significance in the multiple regressor case. In a multiple regression model, first drop all non-constant regressors and consider the regression of y on the sole regressor "x". If this is not significant, then it will obviously remain insignificant after inclusion of additional regressors. If "x" is significant, then regress y on all regressors other than "x". The residuals "y*" from this regression can be considered as the dependent variable "y" purged of the influence of the remaining regressors. Regress these residuals "y*" on "x". If "x" is significant, then it will necessarily be significant in the multiple regression model of y on all the regressors. The case where x is significant when considered as the solitary regressor, but not significant when regressed on y* is ambiguous. This means that the regressor x may or may not be significant depending on the configuration of variances as well as the prioritization of the regressors. Our methods provide a preliminary analysis of significance; a complete analysis will require the extension of our results to the multiple regressor case.

An unsolved puzzle is the invariance conjecture. The maximum bias functions $\mathcal{B}^+(a)$ and $\mathcal{B}^-(a)$ depend directly on the sequence of regressors. Why the value of '$a$' at their intersection depends only on the kurtosis ($K$) is a mystery we leave for more competent mathematicians to resolve.